\documentclass[oneside]{amsart}

\usepackage{amssymb,amsthm,amsmath}
\usepackage{amsfonts,amscd}
\usepackage[mathscr]{eucal}
\usepackage[dvips]{graphicx}

\newcommand{\C}{\mathbb{C}}

\newcommand{\Span}{\mathrm{span}\,}
\newcommand{\bra}{\langle}
\newcommand{\ket}{\rangle}
\newcommand{\pd}{\partial}

\newcommand{\cho}{\begin{pmatrix}}
\newcommand{\ose}{\end{pmatrix}}

\newcommand{\Hol}{H_{0,l}}
\newcommand{\Gol}{G_{0,l}}

\newcommand{\Got}{G_{0,3}}

\newcommand{\Gel}{G_{\epsilon,l}}
\newcommand{\Het}{H_{\epsilon,3}}

\newtheorem{mytheorem}{Theorem}{\bfseries}{\rm}
{\bfseries}{\rm}
{\bfseries}{\rm}
\newtheorem{myproposition}[mytheorem]{Proposition}{\bfseries}{\rm}
{\bfseries}{\rm}
{\bfseries}{\rm}
\newtheorem*{myproposition*}{Proposition}{\bfseries}{\rm}
\newtheorem*{mydefinition*}{Definition}{\bfseries}{\rm}
\newtheorem*{mylemma*}{Lemma}{\bfseries}{\rm}
\newtheorem*{theorem*}{Theorem}{\bfseries}{\rm}

\newcommand{\isotopyfig}{\setlength{\unitlength}{.75cm}
\begin{picture}(5.5,5.5)

\put(2,2){\line(1,0){1}}
\put(3.5,3){\line(-1,-2){.5}}
\put(4,2){$-$}
\put(2.5,1){$-$}

\put(2,4){\line(-1,2){.35}}
\put(3,4){\line(1,2){.35}}
\put(1.5,3){\line(-1,0){.75}}
\put(3,2){\line(1,-2){.35}}

\thicklines

\put(2,2){\line(-1,2){.5}}
\put(2,4){\line(1,0){1}}
\put(2,4){\line(-1,-2){.5}}
\put(3.5,3){\line(-1,2){.5}}
\put(1,2){$+$}
\put(1,4){$+$}
\put(4,4){$+$}
\put(2.5,5){$+$}

\put(2,2){\line(-1,-2){.35}}
\put(3.5,3){\line(1,0){.75}}
\put(4.5,3){$\cdot$}
\put(4.6,3){$\cdot$}
\put(4.7,3){$\cdot$}
\put(1.5,1){$\cdot$}
\put(1.45,.9){$\cdot$}
\put(1.4,.8){$\cdot$}

\thinlines

\put(2.3,3){$-$}
\put(1.75,3){\vector(1,-2){.4}}
\put(2,3.5){\vector(1,-2){.65}}
\put(2.875,3.75){\vector(1,-2){.4}}
\put(2.375,3.75){\vector(1,-2){.65}}
\end{picture}}

\begin{document}
\title{On Freedman's lattice models for topological phases}
\author{James Brink \\Dept of Math., UC Berkeley, jbrink@math.berkeley.edu  \\and Zhenghan
Wang
\\Dept of Math., Indiana University, zhewang@indiana.edu}

\maketitle

\section{Introduction}

The program of topological quantum computation is to realize fault tolerant
quantum computation using topological phases of quantum systems [FKLW].  The central open
question is whether or not there exist such physical systems which
are capable of performing universal quantum computation.  In [F],
 a family of Hamiltonians $H_{0,l}$ is proposed as candidates for the Chern-Simons phases,
 which are known to support
 universal quantum computation for each level $l\geq 3, l\neq 4$ [FLW1,FLW2].
 Freedman conjectures that the perturbed ground states of $H_{0,l}$
are given by the Drinfeld double
of the $SO(3)$-Witten-Chern-Simons topological quantum field theories (TQFTs).
The approach in [F] is an algebraic study of the effect of perturbation based on a rigidity result
of the picture TQFTs [FNWW].  The picture TQFTs are the Drinfeld
double of the $SO(3)$-Witten-Chern-Simons TQFTs.
The idea is to treat local relations in picture TQFTs as perturbations.  In
this paper we will investigate numerically the perturbed ground
states of $H_{0,l}$ for some computationally tractable cases.  While numerical study of Hamiltonians
is carried out routinely in physics literature, we face a dilemma here.
The major motivation of our investigation is
quantum computing, but one motivation of quantum computing as Feynman
pointed out is to study quantum systems numerically.  Therefore,
we are in a self-referential situation.  This is also
manifested in the fact that our numerical computation quickly reaches the limit of
present computing power.

Freedman's Hamiltonians $H_{0,l}$ define quantum loop gas models on any
celluated compact surface.  We study the simplest nontrivial cases: celluations of the
torus.  Our numerical data support Freedman's conjecture, but the conjectured
 space of ground states does not come out in full.  Study of substantially
 larger systems is necessary, but computationally
intractable now.  One new phenomenon
we discovered is some lonely states in the  ground
states of $H_{0,l}$.  Those lonely states give rise to
unwanted ground state vectors of $H_{0,l}$
which persist in the perturbed ground states.
There are several possible explanations of those phenomena:
 the small size of the system, the choice of our
perturbation, or the Euclidean geometry of the torus.
We also observe clearly the expected energy gap
between ground states and the first excited states.

The Hamiltonians $H_{0,l}$ are $7^{\textrm{th}}$ order interaction
Hamiltonians.  It is an open question to find $2^{\textrm{th}}$
order or $3^{\textrm{th}}$ order interaction Hamiltonians with
approximately the same ground states.  Note that
[F] also contains a family of $4^{\textrm{th}}$ order interaction Hamiltonians
similar to $H_{0,l}$.

\section{Freedman's Hamiltonian}

At the heart of this study lie the Hamiltonians $\Hol$, for levels
$l\ge1$, which grants the structure of a TQFT
 to the hypothetical physical systems. Each Hamiltonian $\Hol$ is described as a sum of
local projections, which ``implement'' the concept of combinatorial generalized
isotopy ({\em g(d)-isotopy}) on a celluated surface.

\subsection{Combinatorial Isotopy}
Fix a closed oriented surface $\Sigma$, and let $\Delta$ be a triangulation of $\Sigma$
with $n$ vertices. The dual graph $\Delta^*$ to $\Delta$ therefore
defines a celluation of $\Sigma$ by $n$ polygons, with the edges of
$\Delta^*$ being boundaries of adjacent cells.

A {\em spin configuration} is an assignment $s:\Delta^*\to\{+,-\}$ of
a positive or negative spin to each cell in $\Delta^*$. We denote a
``spin flip'' by $^-$: on entire spin configurations, $\bar{s}$
denotes a global interchange of $+$ and $-$; for a cell $c$ of $\Delta^*$,
$\bar{s}^c$ denotes the
spin configuration which agrees with $s$ away from $c$, and flips
the spin at $c$.

A spin configuration can be thought of as a 2-coloring of $\Sigma$,
partitioning $\Sigma$ into $+$ and $-$ regions. We will be interested in
studying the boundary $\pd_s$ of these $+$ and $-$ regions with
respect to spin configurations $s$.  We will refer to $\pd_s$ as {\em domain
walls}
of a spin configuration.  Since $\Delta^*$ is dual to a triangulation, the
domain walls are all $1$-manifolds in $\Sigma$.  Our goal will
be to capture global properties of domain walls $\pd_s$ in terms of local
manipulations at each cell.

Thus, for a fixed cell $c$, we will need to consider the boundary $\pd
c$ of $c$, which consists of edges in $\Delta^*$. Color each edge on
$\pd c$ according to the neighboring cell; set $\pd_+c$ to be the
edges which bound a $+$-cell, and $\pd_-c$ to be the edges which bound
a $-$-cell. We then say that the pair $(s,c)$ is {\em type-g} if both
$\pd_+c$ and $\pd_-c$ form a connected topological arc. Note that
neither $\pd_+c$ nor $\pd_-c$ can be empty. We define the pair $(s,c)$
to be {\em type-h} if $c$ and all neighboring cells have the same spin
($\pd_{s(c)}c=\pd c$).

The motivation for these definitions is as follows: if $(s,c)$ is
type-g, then the $+$/$-$ boundary $\pd_s$ meets $\pd c$ along one of
the topological arcs $\pd_+c$ or $\pd_-c$ (the one which is opposite
the spin of $c$). By changing the sign of $c$, we replace the part of
$\pd_s$ which meets $\pd c$ with the other of the topological
arcs. But this transformation can be viewed as a fixed endpoint isotopy
of one arc to the other, with isotopy occurring over the 2-cell $c$
(see the figure below). So, if $(s,c)$ is type-g, then $\pd_s$ and
$\pd_{\bar{s}^c}$ are isotopic.

\isotopyfig

For type-h pairs $(s,c)$, the cell $c$ and all of its neighbors have
the same spin. So, if the spin is flipped on cell $c$, then a closed
loop is added to the boundary $\pd_s$. Type-h cells are therefore
responsible for the ``generalized'' aspect of g(d)-isotopy, namely, the
insertion or removal of disc-bounding loops.

Using this terminology, we can consider a combinatorial form of
g(d)-isotopy, relative to the celluation $\Delta^*$. Given a spin
configuration, we can apply: (i) type-g moves, which consist of
flipping the spin of a type-g cell; and (ii) type-h moves, which
consist of flipping the spin of a type-h cell. We say that two spin
configurations are {\em combinatorially g(d)-isotopic} if one can be
reached from the other via a sequence of such moves. Then we have the
following proposition:
\begin{myproposition}\label{iso} {\em Let $s$ and $t$ be combinatorially
g(d)-isotopic spin configurations with respect to the celluation
$\Delta^*$. Then $\pd_s$ is g(d)-isotopic to $\pd_t$.} \qed
\end{myproposition}
The comments above show that this result is immediate. The converse is
not quite true: it is possible to have non-combinatorially g(d)-isotopic
spin configurations $s$ and $t$ such that $\pd_s$ and $\pd_t$ are
g(d)-isotopic. The reason for this is that the tiling could be too coarse
to allow room for g(d)-isotopy to take place; for example, on a hexagonal
tiling of the torus, a spin configuration which assigned different
colors to adjacent vertical rows would have no type-g or type-h cells
(but its boundary would be g(d)-isotopic to that of its dual).

\subsection{Definition of $\Hol$}
We are now ready to consider the definition of the Hamiltonian
$\Hol$. For a triangulation $\Delta$ with $n$ vertices, we
associate the $2^n$-dimensional Hilbert space
$\mathcal{H}=\bigotimes_{i=1}^n \C^2$; let $c_1,\dots,c_n$ denote the
dual 2-cells in $\Delta^*$. Then we can express a basis for
$\mathcal{H}$ by $\{|s\ket\}$, where $s$ runs over spin configurations
and $|s\ket=|s(c_1)\ket\otimes|s(c_2)\ket\otimes\cdots\otimes
|s(c_n)\ket$. Set the parameter $d=2\cos\frac{\pi}{l+2}$, and put:
\[ \Hol = \sum_{\stackrel{(s,c)}{\text{type }g}}
|s-\bar{s}^c\ket\bra s-\bar{s}^c| + \sum_{\stackrel{(s,c)}{\text{type
}h}} |s-\frac{1}{d}\bar{s}^c\ket\bra s-\frac{1}{d}\bar{s}^c|  \]
We interpret these terms as follows: $\Hol$ is ``indifferent'' to
g(d)-isotopy, hence the equal weighting on the type-g terms. The factor
of $\frac{1}{d}$ expresses that ``loops are worth $d$'', as flipping
the type-h configuration creates a closed bounding loop in the spin
configuration.

The primary observation, to be proved in the next section, is that the
ground state $\Gol$ of $\Hol$ consists of equivalence classes of
combinatorially g(d)-isotopic spin configurations.

\section{Theoretical Analysis}
First we consider analytic results relevant to the conjecture. The
ground state of $\Hol$ is determined explicitly.

\subsection{The Ground State $\Gol$}
For a spin configuration $s$, let $n(s)$ denote the number of trivial
(disc-bounding) closed loops in $s$. Thus if $s'$ is the configuration
$s$ with all trivial closed loops removed, we have $d^{n(s)}s'\simeq s$.

\begin{mytheorem} \label{gol} Fix a surface $\Sigma$, a triangulation
$\Delta$ of $\Sigma$, and let $\simeq_\Delta$ be the equivlance
relation of combinatorial g-isotopy on spin configurations with
respect to $\Delta$. Then the ground state $\Gol$ of the corresponding
Hamiltonian $\Hol$ has as a basis:
\[ |v_i\ket = \sum_{s\simeq_\Delta s_i} d^{n(s)}|s\ket \]
where $d=2\cos\frac{\pi}{l+2}$ and the $s_1,\dots,s_k$ are
representatives of the $k$ equivalence classes determined by
$\simeq_\Delta$.
\end{mytheorem}

\begin{proof} First, observe that the subspace $V_i = \Span\{|s\ket :
s\simeq_\Delta s_i\}$ is invariant under $\Hol$; this is immediate from
the definition of $\Hol$, as each projector acts invariantly on one
of the $V_i$ and trivially on $V_j$ for $i\ne j$. So we can write
$\mathcal{H}=\bigoplus_i V_i$, with $\Hol$ acting invariantly on
each summand; therefore, the ground state can be likewise decomposed,
$\Gol= \bigoplus_i \ker(\Hol|_{V_i})$. So the theorem reduces to
the claim that $\ker(\Hol|_{V_i})=\Span|v_i\ket$.

First, we show that $|v_i\ket$ lies in $\ker(\Hol|_{V_i})$. For any
$s\simeq_\Delta s_i$, we have:

\[ \bra s|\Hol|v_i\ket = \sum_{\stackrel{\text{cells $c$ s.t.}}{\text{
  $(s,c)$ type g}}}d^{n(s)}-d^{n(\bar{s}^c)} +
  \sum_{\stackrel{\text{cells $c$ s.t.}}{\text{$(s,c)$ type
  h}}}d^{n(s)}-\frac{1}{d}d^{n(\bar{s}^c)} + \]
\[\hspace{3.5cm}
\sum_{\stackrel{\text{cells $c$ s.t.}}{\text{$(\bar{s}^c,c)$ type g}}}
  d^{n(s)}-d^{n(\bar{s}^c)} + \sum_{\stackrel{\text{cells $c$
  s.t.}}{\text{ $(\bar{s}^c,c)$ type
  h}}}\frac{1}{d^2}d^{n(s)}-\frac{1}{d}d^{n(\bar{s}^c)}\]

For any $(s,c)$ that is type-g, we have $n(s)= n(\bar{s}^c)$, since
the two configurations are (strictly) isotopic. Similarly, if $(s,c)$
is of type-h, then $n(\bar{s}^c)=n(s)+1$, so that
$d^{n(s)}-\frac{1}{d}d^{ n(\bar{s}^c)}=0$; when $(\bar{s}^c,c)$ is
``type h'', then $\frac{1}{d^2}d^{n(s)}-\frac{1}{d}d^{n(\bar{s}^c)}
=0$. So all terms in the summation cancel; since this holds for all
$s\simeq_\Delta s_i$, this shows that $|v_i\ket\in\ker(\Hol|_{V_i})$.

The converse claim follows similarly; for, suppose some nonzero vector
$|v\ket\in V_i$ is in the kernel of $\Hol|_{V_i}$. Then $\bra
s|v\ket\ne0$ for some $s\simeq_\Delta s_i$, so that $\Hol|s\ket\bra
s|v\ket$ will have nonzero components along each $|s'\ket$ which
differs from $s$ by one type-g or type-h move. So in order for
$|v\ket\in\ker(\Hol|_{V_i})$, it must also have nonzero components
along these $|s\ket$. Continuing this argument, we see that $\bra
s|v\ket\ne0$ for all $s\simeq_\Delta s_i$. Analyzing the above
computation, we see that if $s$ and $s'$ differ by a type-g move, then
$\bra s|v\ket=\bra s'|v\ket$; and if they differ by a type-h move
(say, $s'$ is $s$ with a loop removed), then $d\bra s|v\ket=\bra
s'|v\ket$. These requirements in turn force $|v\ket$ to be a multiple
of $|v_i\ket$, so that $\ker(\Hol|_{V_i})=\Span|v_i\ket$.
\end{proof}

\section{Numerical Study}

Celluations of a surface $\Sigma$ dual to triangulations have a nice topological
property.  The domain walls in a spin configuration are all 1-manifolds.
The homology class
represented by those domain walls is the zero class as the domain
walls are bounding 1-manifolds.  The linear combinations of domain
walls coming from different spin configurations in a fixed
celluation forms a finite dimensional vector space.  Those vector
spaces are combinatorial approximations of a picture TQFT.
Local relations in picture TQFTs are the topological realization of
perturbations.
The rigidity of the picture TQFTs says the only non-trivial local
relations for each $l$ is generated by the Jones-Wenzl projector.
Therefore, for each level $l$
perturbations of the
Hamiltonian $H_{0,l}$ will result in only one possibility: the
picture TQFTs at level $l$.
The purpose of the present study then is to perform numerical analysis
on computationally tractable cases, in order to test this conjecture.
We will focus on perturbations of the form
$H_{\epsilon}=H+\epsilon V$.  As a preliminary case, we will choose $V$
to the sum of $\sigma_x$ on each cell.
Our study focuses on $l=3$, so the loop value is $d=\frac{\sqrt{5}+1}{2}$.
This corresponds to the double of
the $SO(3)$-Witten-Chern-Simons theory at level=3. In [FLW1], we have
shown that the level $l=3$ theory for both $SU(2)$ and $SO(3)$ supports universal quantum
computation.

\subsection{On Tilings}

\begin{figure}
\includegraphics[scale=.7]{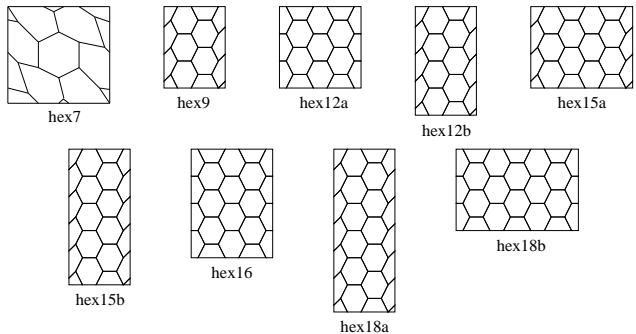}
\caption{The hexagonal tilings of the torus used in the numerical
study.}\label{tilings}
\end{figure}

The numerical study focused on tilings of the torus.
Nine hexagonal tilings of the torus were considered, representing the
``obvious'' hexagonal tilings of the torus which are computationally
tractable. These tilings are depicted in Figure \ref{tilings}, on the
torus visualized in the standard way as a rectangle with opposite
sides identified.

The tiling {\bf hex7} is the dual of a minimal triangulation of the
torus, and thus is a minimal tiling. The other tilings are, loosely
speaking, formed by $p$ adjacent vertical columns of $q$ cells; when
$p$ is odd, an extra ``twist'' is needed to align the vertices
properly. Thus, tilings with
$(p,q)=(3,3),(4,3),(3,4),(5,3),(3,5),(4,4),(3,6),(6,3)$ are
represented.

\subsection{Ground State Vectors of $\Hol$}
First, a combinatorial computation was performed to verify the result
of Theorem \ref{gol}, regarding the form of the ground state vectors
of $\Hol$. Spin configurations were grouped into g-isotopy classes,
and vectors corresponding to each class were created according to the
formula of Theorem \ref{gol}. Each such vector was found to be in the
ground state of $\Hol$; further, the number of such (orthogonal)
vectors was equal to the dimension of the ground state, as calculated
via an eigenvalue computation, so that the ground state was
numerically verified to be exactly the span of such vectors, in all
cases considered.

The interesting result of these calculations concerns what we will
call {\em lonely} configurations: spin configurations which are not
g-isotopic to any other spin configuration. (In other words,
configurations in which there are no type-g or type-h cells.) Some of
the tilings considered admit such lonely configurations, and others do
not; in considering the numerical analysis of the perturbed
Hamiltonians below, it will be necessary to take these lonely
configurations into account.

We considered nine different hexagonal tilings of the torus, ranging
from $7$ cells (the minimal hexagonal tiling of the torus,
corresponding to the minimal triangulation of the torus) to $18$ cells
(the maximum computationally tractable case). The following table
presents the results of this preliminary analysis of the tilings:

\begin{table}
\caption{Ground states of $\Hol$, with respect to given tilings}
\begin{tabular}{|c|c||c|c|c|} \hline
Tiling & $n$ & $\dim\Got$ & Lonely configurations & Non Lonely\\

\hline\hline hex7 & 7 & 5 & 0 &5\\ \hline hex9 & 9 & 5 & 0 &5\\
\hline hex12a & 12 & 8 & 2 &6\\ \hline hex12b & 12 & 17 & 12 &5\\
\hline hex15a & 15 & 7 & 0 &7\\ \hline hex15b & 15 & 8 & 0 &8\\
\hline hex16 & 16 & 24 & 18&6
\\ \hline hex18a & 18 & 16 & 8 &8\\ \hline hex18b & 18 & 21 & 14 &7\\
\hline
\end{tabular}
\end{table}

\subsection{Perturbation Ground States $\Gel$}

For the numerical analysis of the perturbed ground states, we
calculated the lowest energy eigenvalues of $\Het$, for $\epsilon$
ranging from $0$ to $1$. Plots of the eigenvalues as a function of
$\epsilon$ are given below, for each of the nine tilings in the table
above.

As the collected data simply provides the lowest eigenvalues, the
plotted lines simply indicate the trajectories of the various
eigenvalues. The lines themselves do not follow a particular
eigenvalue, but rather, they connect the lowest eigenvalue, the
second-lowest eigenvalue, etc.; in other words, the plots below should
be treated more like scatterplots. The connecting lines are included
to help elucidate the trajectories of the eigenvalues and make it
easier to determine where eigenvalues converge and diverge.

Also, the plots do not indicate the number of eigenvalues represented
by a particular line. For this, the numerical data had to be examined
by hand. Thus all lines represent one eigenvalue, except where
explicitly labelled otherwise. Not all multiplicities in higher energy
eigenstates are labelled.

Finally, the numerical algorithm for calculating eigenvalues did not
always converge, and thus provided possibly erroneous results for
particular values of $\epsilon$. In most cases these were isolated,
and therefore can be safely ignored; however, there are a few larger
regions in which the algorithm failed to converge. These also are
indicated in the plots below, by regions surrounded by dashed lines.

First, we consider the tiling {\bf hex7}:
\begin{figure}[h]
\includegraphics[scale=.7]{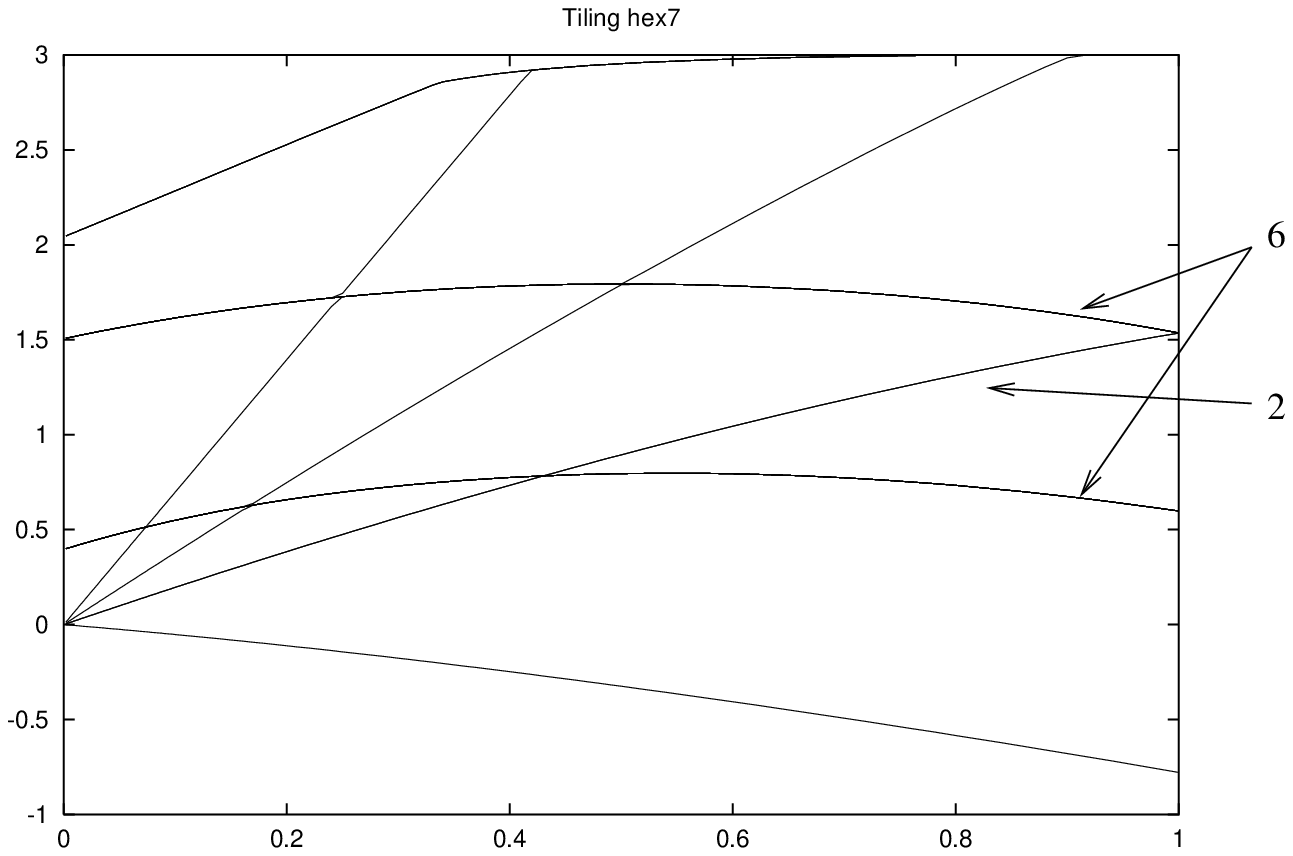}
\end{figure}

The initial 5-dimensional ground state splits into four seperate
states, with the second lowest of these being doubly degenerate, as
indicated in the figure.

For {\bf hex9}:
\begin{figure}[h]
\includegraphics[scale=.7]{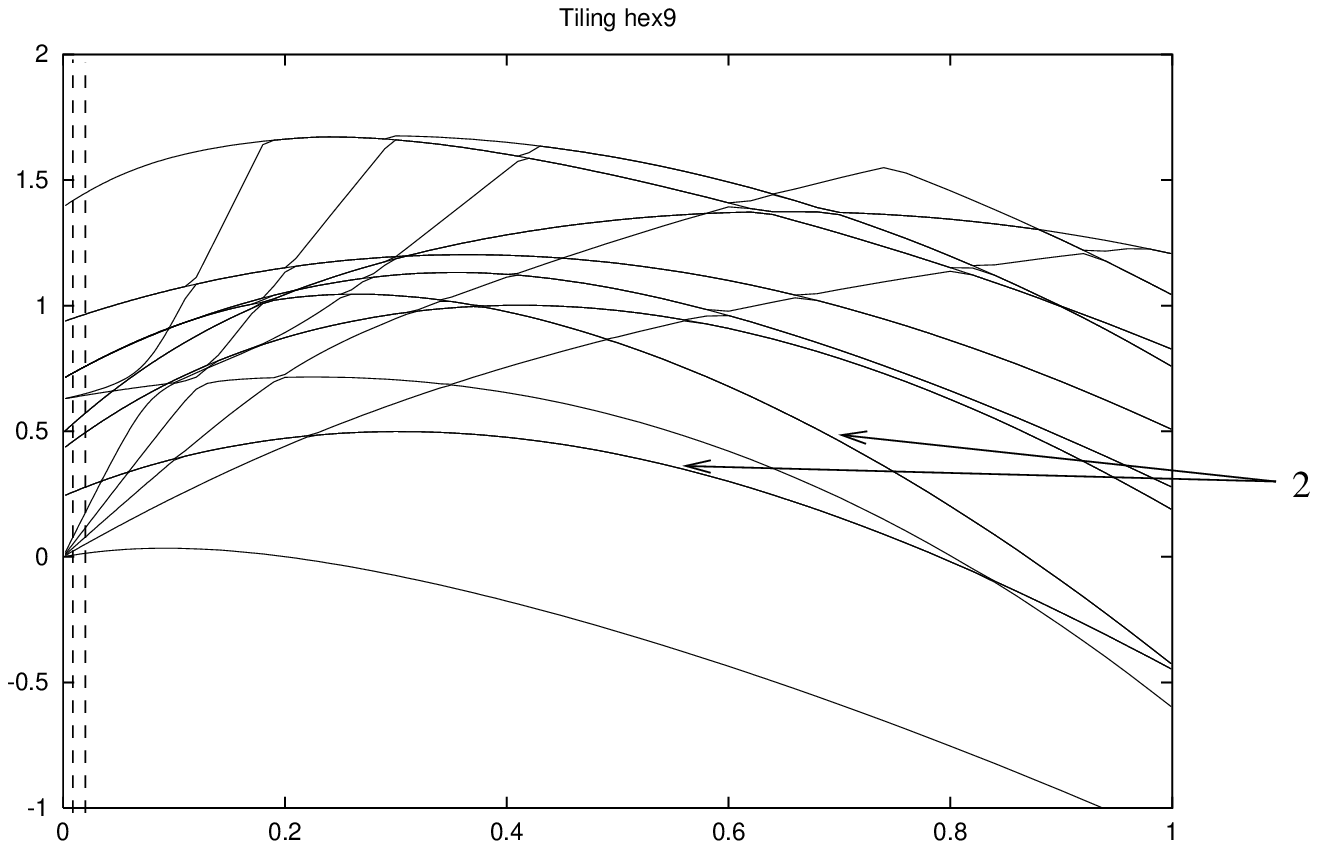}
\end{figure}

As in {\bf hex7}, the initial 5-dimensional ground state splits,
although this time into five seperate states. Again only the lowest of
these remains in the ground state as $\epsilon$ grows. (The indicated
region of failed convergence is for $0.01\leq\epsilon\leq0.02$.)

The spikes appearing in the plot below for {\bf hex12a} are due to
failed convergence of the algorithm; however, the algorithm only
failed to converge for some of the higher energy
eigenvalues. Therefore it is not indicated as a potential error region
(as we are interested mainly in the lowest energy eigenstates).
\begin{figure}[h]
\includegraphics[scale=.7]{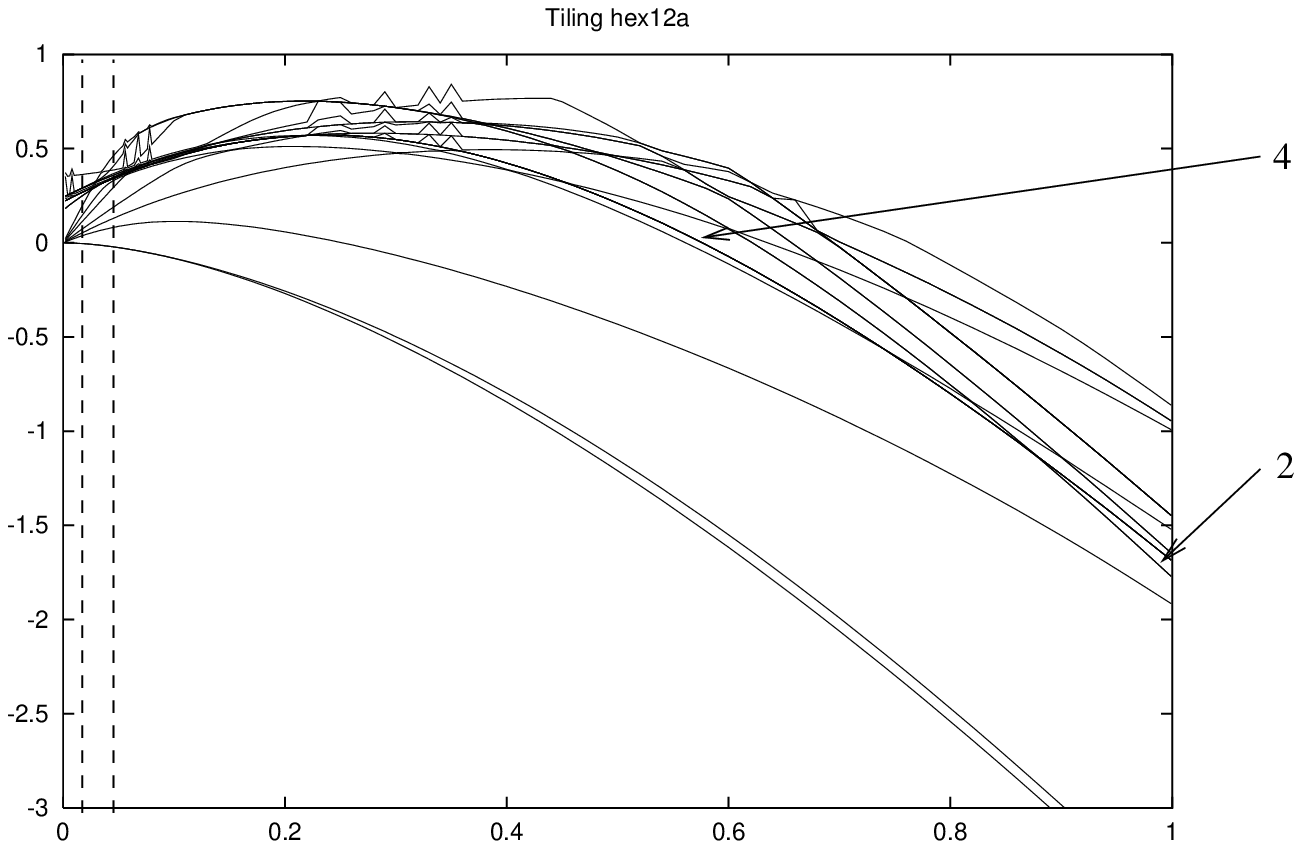}
\end{figure}

Though it is difficult to see, six of the eight original ground state
eigenvalues split in the initial region ($\epsilon\leq0.01$), with the
two lowest staying together. Then the two lower ones fracture apart
around $\epsilon\approx0.2$. Recall also that this tiling has two
lonely configurations, although it is not possible to tell from these
data how the lonely configurations might correspond to the eigenvalue
trajectories.

The tiling {\bf hex12b} has 12 lonely configurations; thus the plot
below indicates more clearly how these relate to the other ground states.
\begin{figure}[h]
\includegraphics[scale=.7]{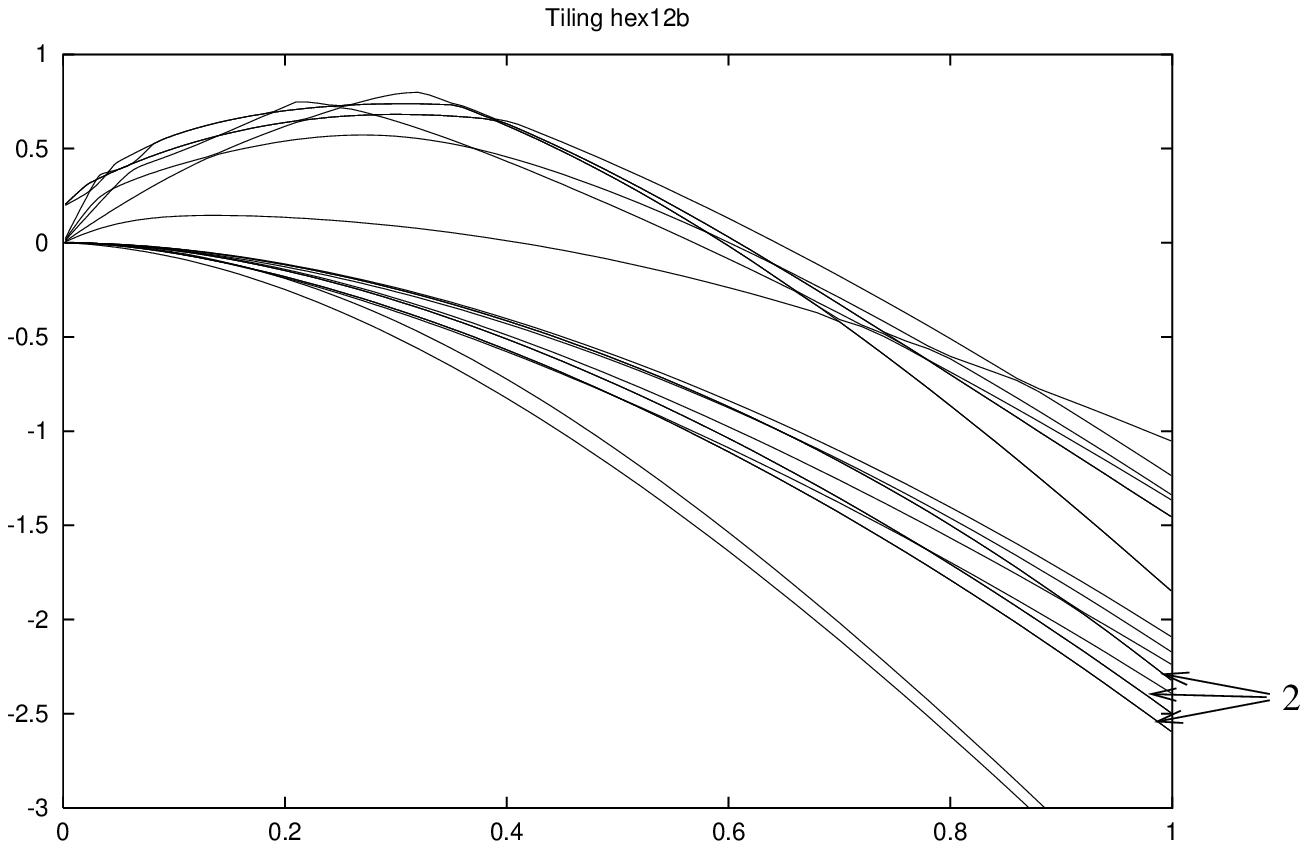}
\end{figure}

From this plot it is almost certain that the 12 eigenvalues that
initially are negative correspond to the 12 lonely configurations of
the tiling.

Neither of the 15-tilings admit any lonely configurations. The plot
for {\bf hex15a} follows:
\begin{figure}[h]
\includegraphics[scale=.7]{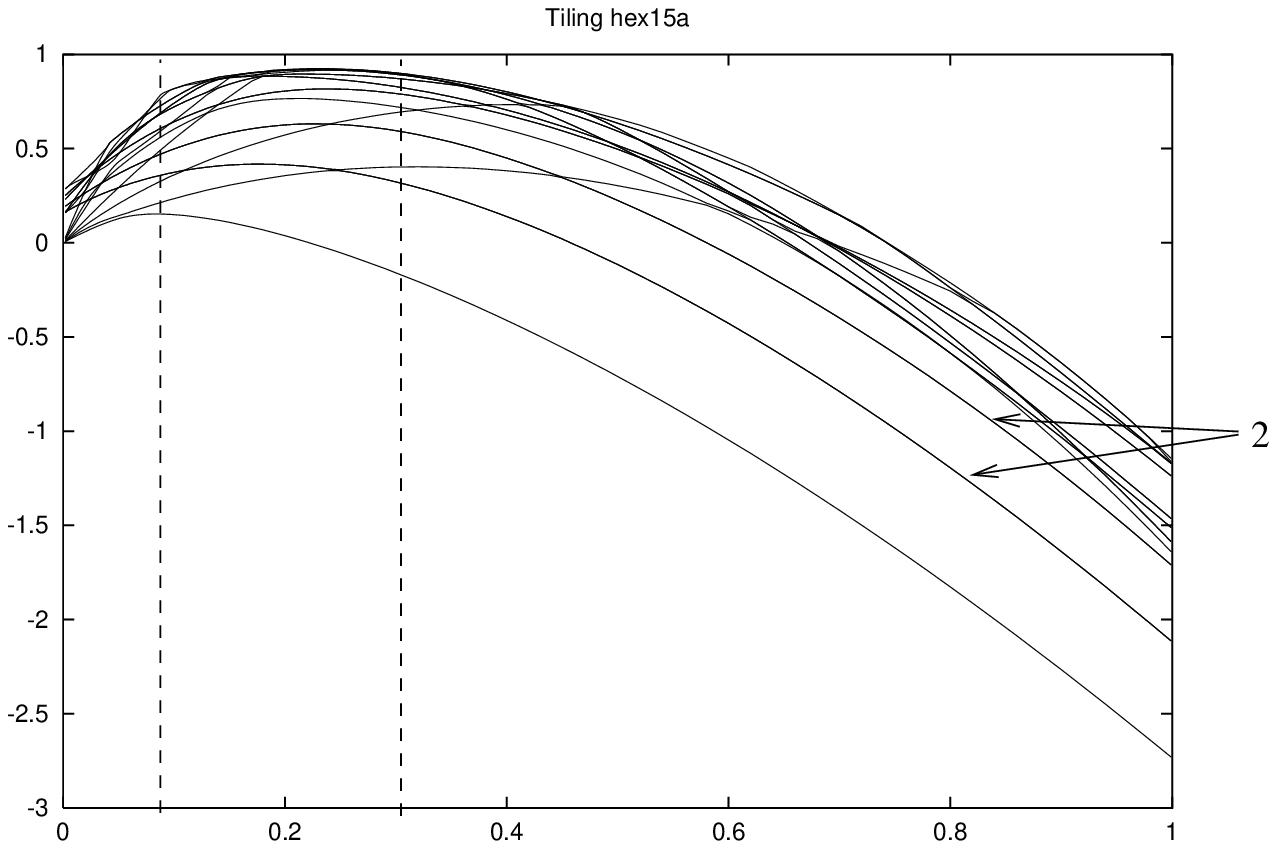}
\end{figure}
All seven of the ground state eigenvalues split, with only one of them
remaining in the ground state for $\epsilon>0.3$. The area of failed
convergence is $0.09\leq\epsilon\leq 0.3$.

For {\bf hex15b}:
\begin{figure}[h]
\includegraphics[scale=.7]{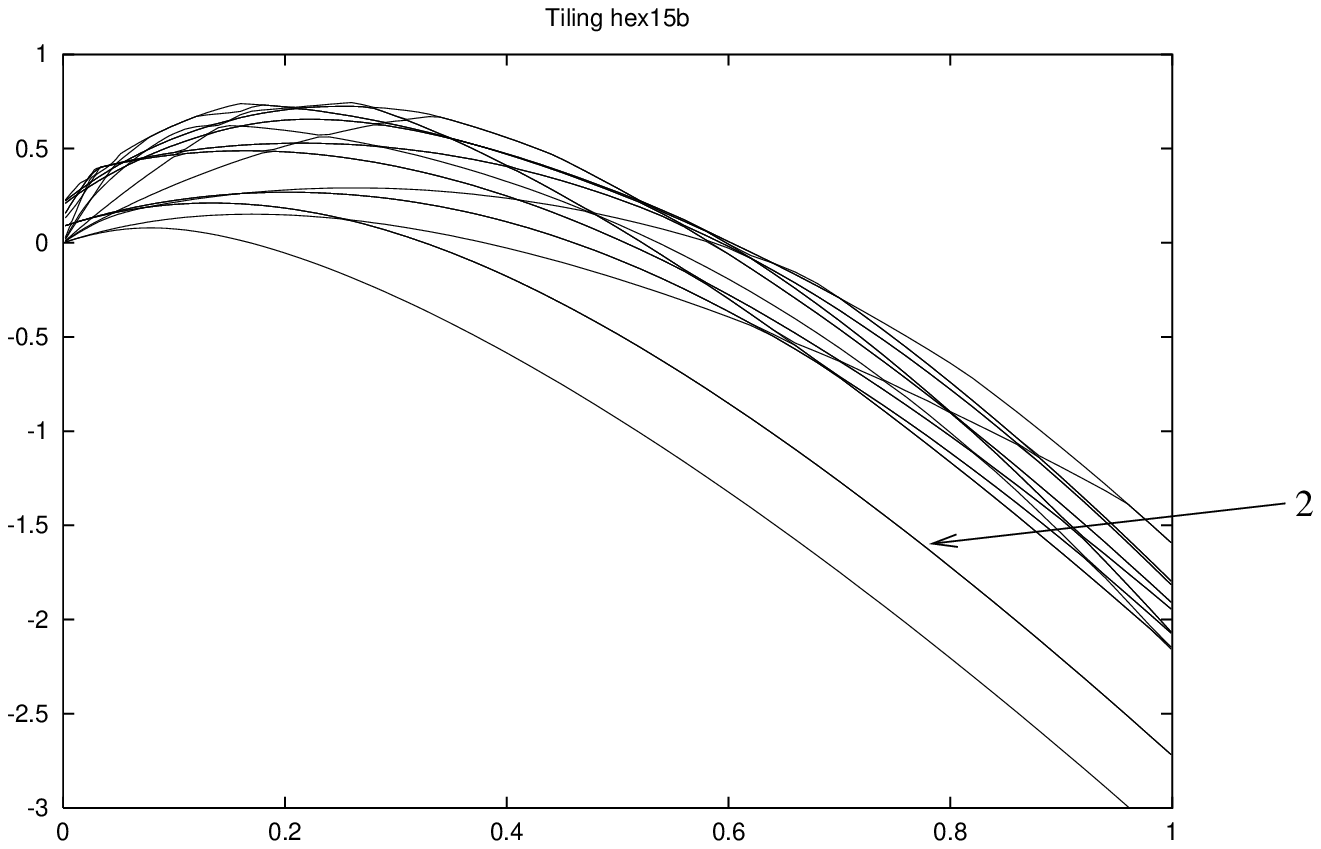}
\end{figure}

Again, all eight of the ground state eigenvalues split, and only one
remains in the ground state beyond $\epsilon\approx0.25$. Though it is
difficult to see on the graph, the line labelled with multiplicity two
is not originally in the ground state when $\epsilon=0$.

\begin{figure}[h]
\includegraphics[scale=.7]{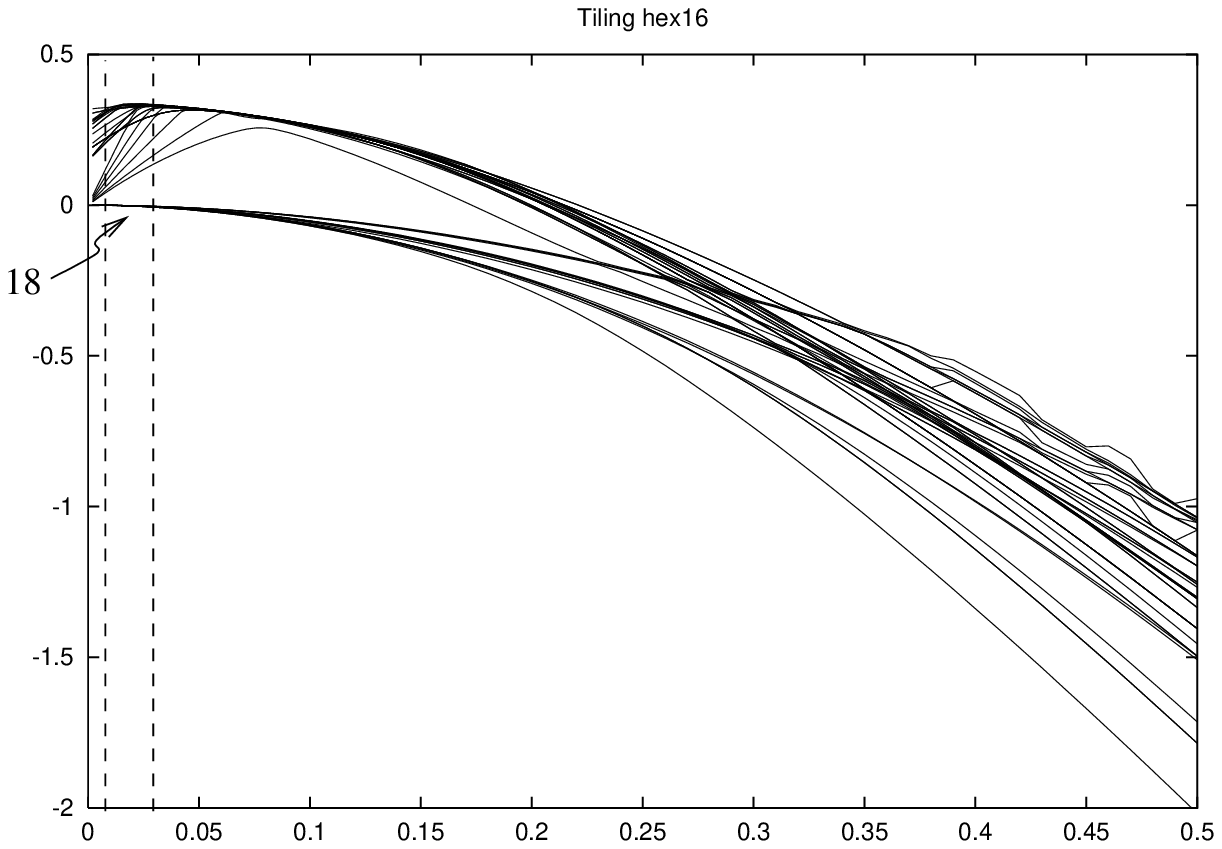}
\end{figure}

The plots for the 16- and 18-tilings are restricted to
$0\leq\epsilon\leq0.5$, as they are computationally expensive, and the
trajectories of the eigenvalues seem clear. For {\bf hex16}, the
general trend regarding the lonely configurations remains consistent;
interestingly, the other pattern seems to hold, although in a slighty
more dramatic fashion. The eigenvalue corresponding to the lowest
non-lonely configuration first rises sharply, then hits a peak around
$\epsilon\approx0.07$, and then begins to decline, remaining below the
other non-ground state trajectories until
$\epsilon\approx0.3$. Although it appears to nearly ``merge'' with the
non-ground state trajectories at $\epsilon\approx0.07$, manual
inspection of the data indicates that in fact it remains clearly
distinct.

The plot for {\bf hex18a} follows:
\begin{figure}[h]
\includegraphics[scale=.7]{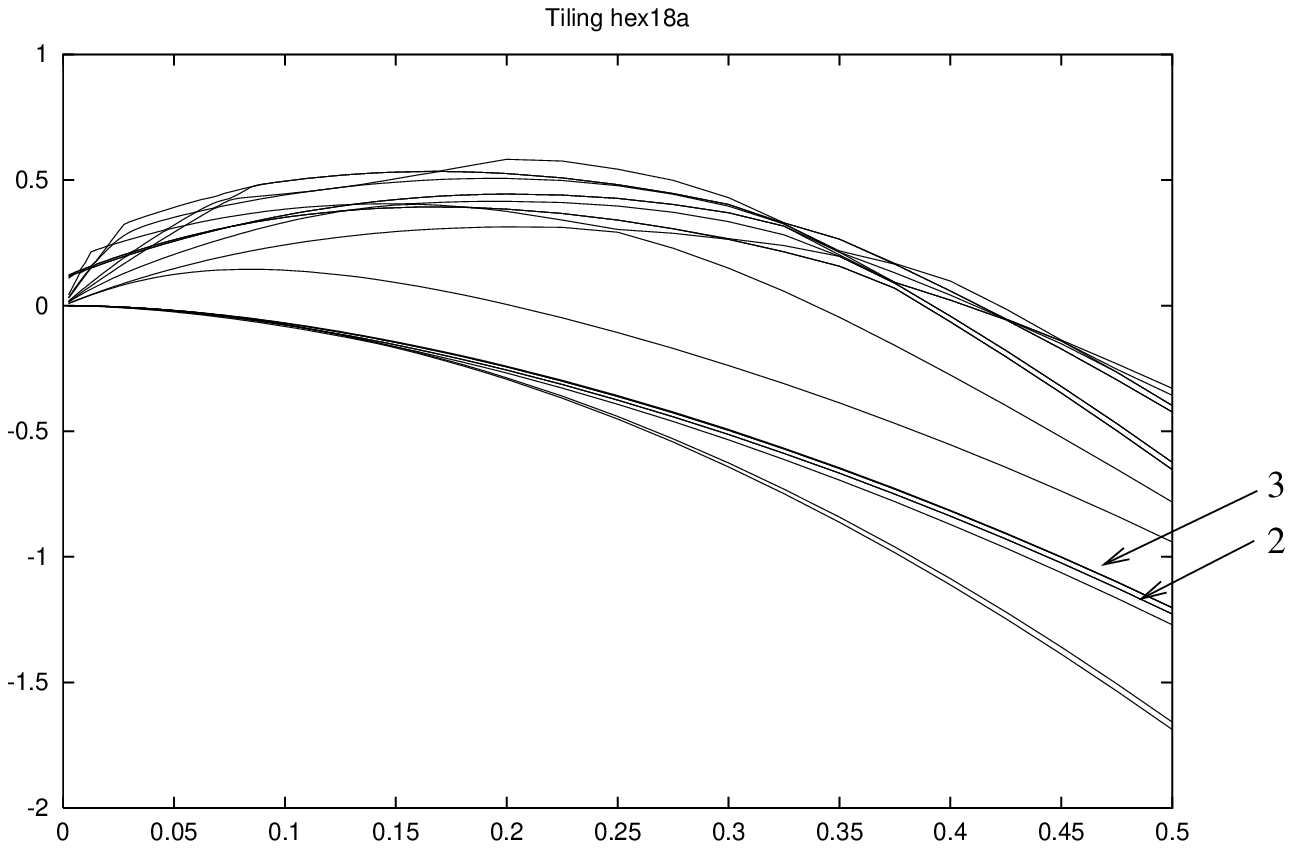}
\end{figure}
Again, eight of the ground state eigenvalues separate distinctly from
the rest, almost certainly corresponding to the eight lonely
configurations. And again, only one of the original ground state
eigenvalues remains consistently low, with only the lonely
configurations below it.

\begin{figure}[h]
\includegraphics[scale=.7]{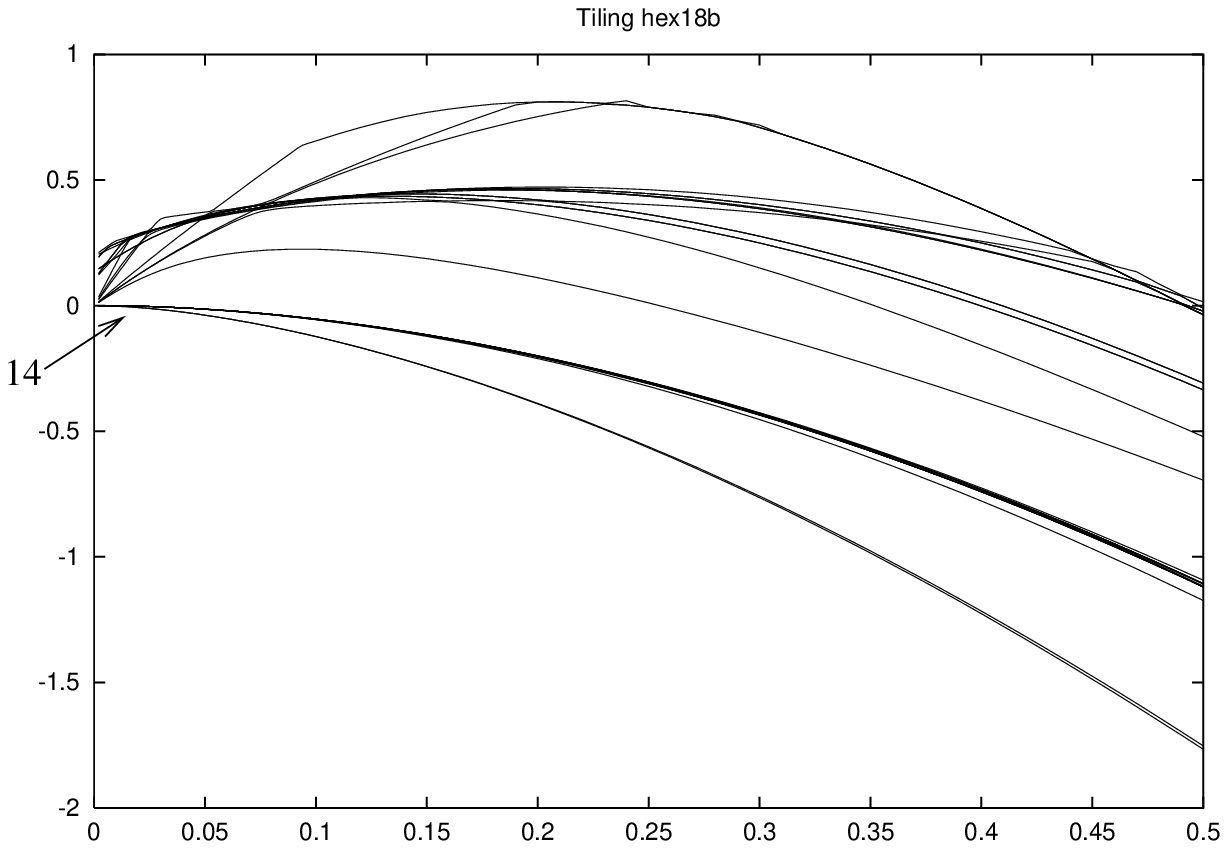}
\end{figure}

Finally, the tiling {\bf hex18b} demonstrates the same pattern. The
fourteen dimensions of the ground state corresponding to the lonely
configurations initially split off and remain consistently lower than
the rest, and one of the remaining ground state configurations
seperates and remains in a lower state than everything but the lonely
configurations.

\section{Discussion}

\begin{itemize}

\item The
expected dimension of the perturbed ground states for $l=3$ is 4.
Ignoring the lonely states represented by lonely configurations,
the un-perturbed ground states form a space of dimensions 5,6,7,
and 8.
There was
consistently one branch of the original ground state which split
off from the rest and remained in the ground state. This is
an indication of what we're looking for, but not coming
out in full because of the small size of the tilings.  Only two of
the tilings ({\bf hex18a} and {\bf hex15b}) even admitted four
parallel essential circles, and even then only in one direction (e.g.,
just horizontally, not vertically).

\item  The problem of lonely configurations might be only a technical
issue.  If they exist, then they
will certainly be the lowest energy eigenstates under perturbation.
It seems likely the lonely configurations are ``isolated'' in the
sense that, if the system is prepared in some suitable initial state,
then it will never ``jump'' to one of these
lonely configurations, but rather slowly fall into one of the ground
state configurations corresponding to a large g-isotopy class.

Given a triangulation of a surface, we can always
subdivide the surface to get a new triangulation without any
lonely configurations.  The problem with this solution is that we
have to introduce much more tilings into the problem, which is definitely
computationally intractable.

\item  Recall that in the Chern-Simons theory, the Lagrange has a quadratic
term $A\wedge \textrm{d}A$ and a cubic term $\frac{2}{3} A^3$.  If the cubic term is
treated as a perturbation, then it
 is a 3rd order term.  So maybe a
better perturbation should be at least of order 3.

\end{itemize}

\section{Gap in thermodynamic limit}

Inspecting the data, we observe that the lowest eigenvalue and the
next one has a clear gap.  The question is whether this gap
persists if we go to finer and finer tilings. Further study will
be carried out.

Acknowledgement:  Research of Z.W. is supported by NSF grant
CISE/EIA-0130388 and Army Research Office.

\appendix

\section{Implementation Details}

The open source numerical package ARPACK \cite{arpack} was the main
tool used for finding the ground state and low energy eigenstates of
the Hamiltonians. ARPACK is optimized to find certain eigenvalues
(e.g., those of the lowest magnitude) of large, sparse, symmetric,
real-valued matrices, and is thus well-suited to our problem. A set of
templates interface the Fortran ARPACK code with C++ was used; all of
the custom code for our problem was done in C++.

For a fixed Hamiltonian, let $n$ denote the number of tiles in the
associated tiling; thus, the dimension of the associated $H_0$ will be
$2^n$.

The sparseness of the Hamiltonian can be analyzed by determining the
number of type-g and type-h spin configuration pairs associated with a
given tiling. It is relatively easy to determine that, for a hexagonal
tiling with $n$ tiles, there are $62\cdot n\cdot2^{n-7}$ type-g and
type-h spin-configuration pairs, and therefore twice as many nonzero
entries in the matrix (total of $2^{2n}$ entries). While this is
relatively sparse, it is still inefficient to compute and store the
matrix explicitly (using the ``Compressed Sparse Column'' format of
ARPACK), as the number of required entries grows
exponentially. Computing the matrix-vector product ``on the fly'' is
much more efficient. This algorithm is given in pseudo-code below:

\begin{verbatim}
//helper method
Bar(c,t); // returns the configuration c with tile t flipped

H0_product(v[],w[]) { //computes H_0|v>=|w>
   w=(0,0,...,0);
   foreach tile t {
      foreach type-h cfg (c,t) {
         w[c]+=v[c]-(1/d)v[Bar(c,t)];
         w[Bar(c,t)]+=(-1/d)v[c]+(1/d^2)v[Bar(c,t)];
      }
      foreach type-g cfg (c,t) {
         w[c]+=v[c]-v[Bar(c,t)];
         w[Bar(c,t)]+=v[Bar(c,t)]-v[c];
      }
   }
}

He_product(v[],w[]) { //computes H_\epsilon|v>=|w>
   H0_product(v,w); //first do the H_0 part
   foreach cfg c {  // then add the perturbation
      foreach tile t {
         w[Bar(c,t)]+=v[c]*epsilon;
      }
   }
}
\end{verbatim}
The algorithm requires $O(n2^{n-7})$ floating-point operations (flops)
per matrix-vector product; compared with the standard product using
the CSC-stored format, which would required $O(2^{2n+1})$ flops. The
configurations that are type-h and type-g, with respect to each
tile, are calculated beforehand and stored. Therefore, the storage
requirements are $62n2^{n-7}$ integers, which is on the same order as
the storage requirements for storing the matrix in CSC format.

An attempt was made to parallelize the code, as there is a parallel
version of ARPACK available. However, the code only allows for
parallelization of the matrix-vector product; and the lag in
communicating results across nodes would be significantly more costly
than the actual computation of the results. Thus the only way in which
parallelization might be useful is if the actual eigenvalue algorithm
was parallelized. This approach was not pursued.

\end{document}